\documentstyle[preprint,eqsecnum,aps,epsfig]{revtex}

\def\({\c c}
\def\|{\'\i}

\newcommand{\bra}[1]{\left\langle{#1}\,\right|\,}
\newcommand{\ket}[1]{\,\left|\,{#1}\right\rangle}

\def\ra{\rightarrow}

\def\lapp{\hbox{$ {     \lower.40ex\hbox{$<$}
                   \atop \raise.20ex\hbox{$\sim$}
                   }     $}  }
\def\rapp{\hbox{$ {     \lower.40ex\hbox{$>$}
                   \atop \raise.20ex\hbox{$\sim$}
                   }     $}  }
\def\be{\begin{equation}}
\def\ee{\end{equation}}
\def\bea{\begin{eqnarray}}
\def\eea{\end{eqnarray}}
\def\non{\nonumber}
\def\bb{\bibitem}

\begin{document}
\draft
\tighten
\preprint{\vbox{\hfill SLAC--PUB--8961\\
\null\hfill UK/TP 01--05\\
\null\hfill August 2001\\}}

\title{Evading the CKM Hierarchy: Intrinsic Charm 
in B Decays\thanks{Research partially
supported by the Department of Energy under
contracts DE-AC03-76SF00515 and DE-FG02-96ER40989.}}

\author{S.~J. Brodsky\thanks{E-mail:\quad sjbth@slac.stanford.edu}}
\address{Stanford Linear Accelerator Center,
Stanford University\\ Stanford, California 94309}

\author{S. Gardner\thanks{E-mail:\quad gardner@pa.uky.edu}}

\address{Department of Physics and Astronomy,
University of Kentucky\\ Lexington, Kentucky 40506-0055}

%\date{August 14, 2001}
\maketitle

\begin{abstract}
We show that the presence of intrinsic charm
in the hadrons' light-cone wave functions,
even at a few percent level,
provides new, competitive decay mechanisms for $B$ decays
which are nominally
CKM-suppressed. For example,
the weak decays of the $B$-meson to
two-body exclusive states consisting of strange plus light
hadrons,
such as $B \to \pi K$, are expected to be
dominated by penguin 
contributions since the tree-level $b\to s
u{\overline u}$ decay is CKM suppressed.
However,
higher Fock states in the $B$
wave function containing charm quark pairs
can mediate the decay
via a CKM-favored
$b\to s c{\overline c}$ tree-level transition.
Such intrinsic charm contributions
can be phenomenologically significant. Since they mimic the
amplitude structure of ``charming'' penguin contributions,
charming penguins need not be penguins at all.
\end{abstract}

\newpage

\section{Introduction}
\label{intrO}

It is usually assumed in the analysis of
$B$-meson decays that
only the valence quarks of the initial and
final-state hadrons participate in the weak interaction.
Typical examples are the semi-leptonic decay $B^- \to \ell^- \bar
\nu \pi^+$, which is based on the transition
$b \to u \ell^- \bar \nu$;
$B^- \to K^- \gamma$, which is based on the penguin amplitude $b \to s
\gamma$; and $B^- \to K^-\pi^0$, which is based on 
$b \to s u  \bar u$ and penguin
$b \to s g^* \to s u \bar u $ transitions. In each case,
it is assumed that the matrix elements of the operators
of the  effective weak Hamiltonian involve only the valence quarks of
the incoming and outgoing hadrons.  Any non-valence gluon or sea quarks
present in the
initial or final state wave functions appear only as spectators.

The wave functions of a bound state in a relativistic quantum field
theory such as QCD necessarily contain Fock states of arbitrarily high
particle number.  For example, the $B^-$-meson has a Fock state
decomposition
\begin{equation}
 \ket{B^- } =
\psi_{b \bar u} \ket{ b \bar u }+
\psi_{b \bar u g} \ket{ b \bar u g }+
\psi_{b \bar u d \bar d} \ket{ b \bar u d \bar d } +
\psi_{b \bar u s \bar s} \ket{ b \bar u s \bar s } +
\psi_{b \bar u c \bar c} \ket{ b \bar u c \bar c } + \cdots.
\end{equation}
The Fock state decomposition is
most conveniently done at equal light-cone time
$\tau = t + {z/ c}$ using light-cone quantization in the
light-cone gauge $A^+ =0$~\cite{Dirac:1949cp,Brodsky:1989pv}.
The light-cone wave function $\psi_n(x_i, \vec{k}_{\perp i},
\lambda_i)$ depends on the momentum fraction of parton $x_i$,
where $x_i = {k^+_i/ P^+}$ and $\sum_i x_i=1$, the
transverse momentum $\vec{k}_{\perp i}$
where $\sum_i \vec{k}_{\perp i}=0$,
and the helicity $\lambda_i$. The light-cone wave functions are
Lorentz invariant; i.e., they are independent of the total momentum
$P^+ = P^0 + P^z$ and of $P_\perp$ of the bound state.
The extra gluons and quark pairs in the
higher Fock states arise from the QCD interactions.  Contributions which
are due to a single gluon splitting such as $g \to c \bar c$ are
associated with DGLAP evolution,
or they provide perturbative
loop corrections to the operators; they are extrinsic to the
bound-state nature of the hadron.  In contrast, the $ c\bar c$ pairs
which  are multiply-connected to the valence quarks  cannot be
attributed to the gluon substructure and are intrinsic to the
hadron's structure. The intrinsic, heavy quarks are thus part of the
non-perturbative
bound state structure of the hadrons themselves~\cite{Brodsky:1980pb},
rather than
part of the short-distance
operators associated with the DGLAP evolution of structure functions or
radiative corrections to the effective weak Hamiltonian.

Recently Franz, Polyakov, and Goeke
have analyzed the properties of the intrinsic heavy-quark
fluctuations in hadrons using the operator-product
expansion~\cite{Franz:2000ee}.
For example, the light-cone
momentum fraction carried by intrinsic heavy
quarks in the proton
$x_{Q \bar Q}$ as measured by the $T^{+ + }$ component of the
energy-momentum tensor is related in the heavy-quark limit to the forward
matrix element $\langle p \vert {\hbox{tr}_c}
{(G^{+\alpha} G^{+ \beta} G_{\alpha
\beta})/ m_Q^2 }\vert p \rangle ,$ where $G^{\mu \nu}$ is the gauge
field strength tensor. Diagrammatically, this
can be described as a heavy quark loop in the
proton self-energy with four gluons
attached to the light, valence quarks~\cite{Brodsky:1984nx,Brodsky:1984sa}.
Since the non-Abelian commutator $[A_\alpha, A_\beta]$ is involved,
the heavy quark pairs in the proton wavefunction are necessarily in a
color-octet state.
It follows from dimensional analysis that the momentum fraction carried by
the $Q\bar Q$ pair scales as $k^2_\perp / m^2_Q$ where
$k_\perp$ is the typical momentum in the hadron wave function.  In
contrast, in the case of Abelian theories,
the contribution of an intrinsic, heavy lepton pair
to the bound state's structure first appears in
${\cal O}(1/m_L^4)$. One relevant operator corresponds
to the Born-Infeld $(F_{\mu\nu})^4$
light-by-light scattering insertion, and the
momentum fraction of heavy leptons in an atom scales as
$k^4_\perp / m_L^4$.

In the case of the proton,
analyses~\cite{Hoffmann:1983ah,Harris:1996jx,Steffens:1999hx}
of the charm structure function
measured by the EMC group
indicates a significant charm quark excess beyond DGLAP or
gluon-splitting predictions
at large $x_{Bj} \sim 0.4,$ and
suggest that the intrinsic charm (IC)
probability is $\stackrel{<}{\sim} 1\%$.
Although these analyses are not conclusive~\cite{Amundson:2000vg},
this value is consistent with the theoretical estimate of
Franz et al.~\cite{Franz:2000ee,comment}
An intrinsic charm component
in the light
hadrons of this scale has been invoked to explain the ``$\rho\pi$'' puzzle
in J/$\psi$ decay~\cite{Brodsky:1997fj}, leading charm production in
$\pi\,N$ collisions~\cite{Brodsky:1992dj,Vogt:1992ki,Vogt:1995zf},
as well as
the production of pairs of  $J/\psi$ at large $x_F$ in these
reactions~\cite{Vogt:1995tf}.

The existence of intrinsic charm (IC) in the proton also implies
the existence of IC in other hadrons, including
the $B$-meson. In order to translate the estimate of IC probability in the
proton  to the IC of a $B$-meson, we
are faced with two conflicting effects. The typical internal
transverse momentum $k_\perp$
is larger in the B-meson, evidently favoring a larger IC
probability in the B meson; on the other hand
the proton's additional valence quark generates
a larger combinatoric number of IC diagrams,
favoring a larger IC probability
in the proton.
In evaluating the first effect, it is useful to compare
positronium with H-atom: the kinematics of the heavy-light system
make its ground-state radius a factor of two
smaller than that of positronium,
and thus its typical bound state momentum is a factor of
two larger. This analogy should be applicable
when comparing the internal scales of the B-meson to that of
the light pseudoscalars; we note that the
normalization of the light-cone wave function
$\phi_i(x,\vec{k_\perp})$ with $i\in \pi,B$ is set by
the decay constant $f_i$. Lattice calculations indicate
$f_B\sim 191$ MeV~\cite{Bernard:2001ht}, so that
$f_B/f_\pi \sim 2$, suggesting that
the momentum $k_\perp$ is significantly higher in the
$B$ meson than in light hadrons.
Thus the IC 
component in the B-meson could be as large as four times that of
the proton, that is, $\sim 4\%.$ The IC component of the $\Lambda_b$
baryon could be larger; in this case, the additional valence
quark generates a larger combinatoric number of IC diagrams as well.
The ways in which the decaying $b$ quark interacts with its
hadronic environment, particularly ``spectator effects,'' 
are evidently 
important in explaining the lifetime difference in the $B$
and $\Lambda_b$~\cite{Neubert:1997we,Voloshin:2000ax} hadrons; 
IC could play a role in this context as well.

The presence of intrinsic charm quarks in the $B$ wave function provides
new mechanisms for $B$ decays.  For example, Chang and Hou have considered
the production of final states with three charmed quarks such as 
$\bar B \to
J/\psi D \pi$ and $\bar B \to J/\psi D^*$~\cite{Chang:2001iy};
these final states are difficult
to realize in the valence model,
yet they occur naturally when the
$b$ quark of the intrinsic charm Fock state $\ket{ b \bar u c \bar c}$
decays via $b \to c \bar u d$.  In fact, the $J/\psi$ spectrum for
inclusive $B \to J/\psi X$ decays measured
by CLEO and Belle shows a distinct
enhancement at the low $J/\psi$ momentum where such decays would
kinematically occur~\cite{Balest:1995jf,Schrenk}.  
Alternatively, this excess could
reflect the opening of baryonic channels such as 
$B^- \to J/\psi \bar p
\Lambda$~\cite{Brodsky:1997yr}.

\begin{figure}
\begin{center}
\vspace{2cm}
\epsfig{file=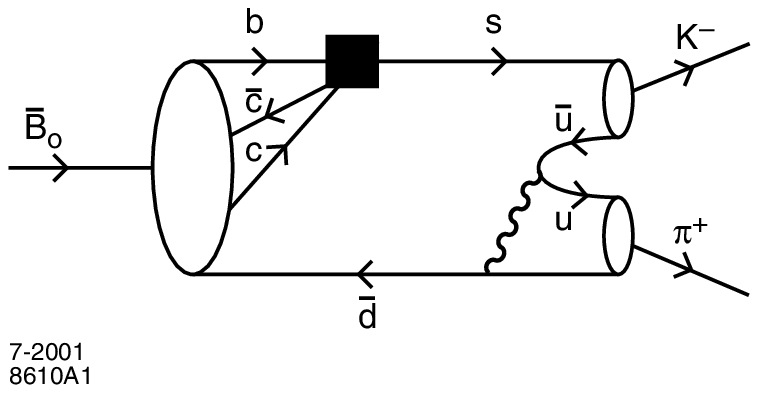, width = 9cm}
\end{center}
\caption{
Intrinsic charm in the B-meson can mediate the decay to a
strange, charmless final state via the weak transition
$b\to s c {\overline c}$. The square box denotes the
weak transition operator.
}
\label{fig1}
\end{figure}

These ideas take on particular significance in view of
the hierarchical structure of the CKM matrix
--- the
weak transition $b\ra s c{\overline c}$ is doubly Cabibbo
enhanced with respect to a $b\ra s u{\overline u}$ transition.
For example, intrinsic charm components in the initial and final
hadron light-cone wave functions
will allow $|\Delta S|=1$ B-meson decays through processes
such as that shown in Fig.~\ref{fig1};
the small intrinsic charm probability
is offset by the comparatively large CKM matrix
elements associated with the $b\to s c {\overline c}$ transition,
promoting their phenomenological
impact.

As a specific illustration,
consider  the exclusive $|\Delta S|=1$ decays,
$B\to \pi K$ and $B\to \rho K$. The various $\pi K$
final states from ${\bar B}^0$ and $B^-$ decay are connected by
isospin symmetry; the same is true of the branching ratios
to $\rho K$. The presence of weak transitions involving
intrinsic charm can alter
the pattern of the predicted branching ratios.
Since the same
initial and final states are involved, the intrinsic charm
contribution  can
interfere with the conventional amplitudes, and yield significant
effects.
We note that such intrinsic contributions function
in a manner identical to that of charm-quark-mediated 
penguin contributions~\cite{Colangelo:1989gi,Buras:1994pb} --- termed 
 ``charming penguins'' 
by Ciuchini et al.~\cite{Ciuchini:1997hb,Ciuchini:1998rj,Ciuchini:2001gv}
--- so that charming penguins need not
be penguins at all.

Halperin and Zhitnitsky have considered the role of
IC in mediating the decays $B\to \eta^\prime K$~\cite{Halperin:1997as},
and $B\to \eta^\prime X$~\cite{Halperin:1998ma}, arguing, as
we have, that IC can be important
when coupled with the Cabibbo-enhanced $b\to s c{\bar c}$ transition
in decays to charmless final states~\cite{Petrov:1998yf}.
They effect their numerical
estimates in the factorization approximation, so that the importance
of their IC mechanism is determined by the parameter
$f_{\eta^\prime}^{(c)}$, where
\begin{equation}
\langle 0 | {\bar c}\gamma_\mu \gamma_5 c| \eta^\prime(p)\rangle
= i f_{\eta^\prime}^{(c)} p_\mu \;.
\end{equation}
Recent work has shown
$f_{\eta^\prime}^{(c)}$ to be $\approx -2$ MeV~\cite{Franz:2000ee},
rather smaller~\cite{Yuan:1997ts,Ali:1998ex,Petrov:1998yf,Feldmann:1998vh}
than
$f_{\eta^\prime}^{(c)}\approx 50-180$
MeV~\cite{Halperin:1997as,Halperin:1998ma},
so that efforts to reconcile the observed
rate with Standard Model (SM) predictions continue~\cite{Kou:2001pm}.
Although other mechanisms could well
be at work~\cite{Hou:1998wy,Kagan:1997qn},
we wish to point out that the factorization approximation
does not capture the physics of IC.
IC is produced in a higher Fock component of a hadron's light-cone
wave function; it is naturally in a color octet state~\cite{Yuan:1997ts},
so that the dynamical
role it plays in mediating B-meson decay is intrinsically
non-factorizable in nature.

Although we will specifically consider the role of IC in exclusive B-meson
decays in this paper,  the effect of IC
can have a more general phenomenological impact on B physics.
For example,  it is well-known that the semileptonic branching
fraction in inclusive B-meson decay, B$_{\rm sl}$, is smaller than
SM predictions; however, the
``natural'' resolution of this puzzle -- an increased
$b\to s c{\bar c}$ rate --  is at odds with the observed
number of charm (and anti-charm) quarks per B-meson decay, $n_c$, as
this is also too small compared to SM expectations~\cite{Falk:1995hm}.
IC in the B-meson can increase the charmless $b\to s$
rate, as illustrated in Fig.~\ref{fig1},
thus reducing the semileptonic branching ratio.
Earlier work ascribed a possible role to IC in resolving
the $B_{\rm sl}/n_c$ puzzle~\cite{Dunietz:1998zn}, yet only
IC in the light hadrons was considered. We believe that
the role played by IC in the B meson in realizing strange,
charmless final states to be potentially of greater importance.
It should be recognized that
IC does not significantly increase the inclusive yield of
charmed hadrons, since the materialization of
intrinsic charm is dynamically suppressed. For example, in
hadron collisions the probability of materializing IC Fock states
is of ${\cal O}(1/m_c^4)$~\cite{Brodsky:1984nx},
save for an exceptional portion of phase space, at large
$x_F$~\cite{Brodsky:1992dj}, for which it is of
${\cal O}(1/m_c^2)$~\cite{Brodsky:1992dj}.
As noted by Chang and Hou,~\cite{Chang:2001iy}
the materialization of IC of the B-meson will
lead to novel exclusive decays to charm final states;
for example, IC can mediate
$B^-\to J/\psi e^- {\bar \nu}_e$, as well as
$B\to J/\psi \gamma$. If there were no IC in the B-meson, such final
states could only be realized through OZI-violating processes.

\section{IC in $B\to \pi K$ and $B\to\rho K$ Decay}

We will now consider the specific role of IC in mediating the exclusive
decays $B\to \pi K$ and $B\to \rho K$.
The operator-product
expansion and renormalization group methods provide a systematic
theoretical framework for analyzing exclusive hadronic B-meson decays.
The amplitude for the decay of a $B$ meson to a hadronic final
state $f$ is given by $A(B\to f) = \bra{f}{\cal H}_{\rm eff}\ket{B}$,
and for $b\to s q{\bar q}$ decay ${\cal H}_{\rm eff}$
can be written as
\be
{\cal H}_{\rm eff} = \frac{G_F}{\sqrt{2}}
\left[
\sum_{p=u,c} V_{pb}V_{ps}^\ast(C_1{O}_1^{p} + C_2{O}_2^{p})
- V_{tb}V_{ts}^\ast \sum_{j} C_j {O}_j
\label{heff}
\right]\;,
\ee
where the  explicit form of the operators are given in Ref.~\cite{lks}.
The $O_{1,2}^{p}$ are the left-handed current-current
operators arising from W-boson exchange; the sum over $j$
contains the strong and electroweak penguin
operators. The Wilson
coefficients are computed at $\mu= M_W$
and are evolved, using renormalization group methods,
to $\mu = {\cal O}(m_b)$; only
$C_2$ is of ${\cal O}(1)$ at the scale $\mu \sim M_W$.
The $C_i(\mu)$ are known; the operator matrix elements
$\bra{f}O_i(\mu)\ket{B}$, however, pose a continuing
theoretical challenge.
Various methods exist to estimate
them~\cite{BSW,Ali:1998nh,Chen:1999nx,hnli,Li:1995zm,Beneke:1999br};
however, all of these approaches
assume that only the valence
degrees of freedom participate in the decay process.
Indeed, this viewpoint is shared by attempts
to catalog all the possible contributions
to the various exclusive decay amplitudes~\cite{Buras:2000ra}.
It is this assumption that we challenge.

It should be emphasized that Fock states of arbitrary
particle number are necessary to describe a relativistic  bound state.
Furthermore, as shown in Ref.~\cite{Brodsky:1999hn}, the matrix element
associated with a time-like form factor entering semi-leptonic
decay has two distinct contributions: a contribution in which
parton number is conserved, so that
$n\; \hbox{partons} \to n \; \hbox{partons}$, and one in
which it is not.
In the latter case, one of the anti-quarks
in a non-valence Fock state fluctuation in the B-meson
annihilates the $b$ quark, yielding the transition
$n + 2 \; \hbox{partons} \to n \; \hbox{partons}$.
These conclusions emerge from Lorentz invariance in concert
with the kinematic constraint of $q^+ >0$.
The omission of such contributions can become acute
in the context
of $b\to s q \bar q$ decays,
since CKM factors favor
$b \to s t\bar t$ and $b\to s c\bar c$  transitions
over those of  $b \to  s u\bar u$ by a numerical factor of
roughly 50.
We shall show that when such decays
are mediated by IC
contributions, the CKM hierarchy can be evaded, impacting
not only the  branching ratios, but also the CP asymmetries associated
with these decays.

Consider the following family of decay modes:
\begin{equation}
{\cal B}({\bar B}^0 \to \pi^+ K^-)
\quad
{\cal B}({\bar B}^- \to \pi^0 K^-)
\quad
{\cal B}(B^- \to \pi^- {\bar K}^0)
\quad
{\cal B}({\bar B}^0 \to \pi^0 {\bar K}^0) \;,
\quad
\end{equation}
as well as
\begin{equation}
{\cal B}({\bar B}^0 \to \rho^+ K^-)
\quad
{\cal B}({\bar B}^- \to \rho^0 K^-)
\quad
{\cal B}(B^- \to \rho^- {\bar K}^0)
\quad
{\cal B}({\bar B}^0 \to \rho^0 {\bar K}^0) \;.
\quad
\end{equation}
These decays are mediated by the short-distance
weak transition $b\to s q\bar q$ where $q\in u,c,t$.
The magnitude of the branching ratios themselves
as well as the precise patterns of ratios of
ratios are sensitive to the contributing decay topologies,
the CKM factors accompanying a particular operator, and the
numerical values of the accompanying Wilson
coefficients. Intrinsic charm can play a particularly
important role here since the operator with
the largest Wilson coefficient combined with the largest combination
of CKM matrix elements
can now enter (the $p=c$ term in Eq.~(\ref{heff})) at tree level.

Let us begin by a constructing a simple numerical
estimate for the above ratios. The transitions
$b \to s t\bar t$ and $b\to s c\bar c$  are favored
over  the $b \to  s u\bar u$ transition by a factor of $\lambda^2$;
however, the
values of the CKM parameters can
exacerbate the flavor hierarchy.
That is, using $\sqrt{\rho^2 + \eta^2}=0.38$ and $\lambda=0.2196$,
which fall within the $\ge 5\%$ C.L. of recent fits~\cite{Hocker:2001xe},
the $b \to  s u\bar u$ transition is suppressed with respect
to $b \to s t\bar t$ and $b\to s c\bar c$ transitions
by a factor of roughly 50. If we neglect isospin-violating
effects, in particular
$|\Delta I|=1$ electroweak penguins, and assume that
$b \to  s u\bar u$ transitions are negligible, just
one decay topology exists for each decay in a particular
``family,'' and the various decay rates are related
by isospin symmetry.
This simple model predicts
the following pattern of ratios of branching ratios:
\begin{equation}
{\cal B}({\bar B}^0 \to \pi^+ K^-) :
{\cal B}({\bar B}^- \to \pi^0 K^-) :
{\cal B}(B^- \to \pi^- {\bar K}^0) :
{\cal B}({\bar B}^0 \to \pi^0 {\bar K}^0)
= 2 : 1 : 2 : 1 \;,
\label{naive}
\end{equation}
as well as
\begin{equation}
{\cal B}({\bar B}^0 \to \rho^+ K^-) :
{\cal B}({\bar B}^- \to \rho^0 K^-) :
{\cal B}(B^- \to \rho^- {\bar K}^0) :
{\cal B}({\bar B}^0 \to \rho^0 {\bar K}^0)
= 2 : 1 : 2 : 1 \;.
\label{naive1}
\end{equation}
[The factors of 2 arise as the
contributing components of the $\rho^\pm$, $\pi^\pm$
and $\rho^0$, $\pi^0$ wave functions
differ by a normalization factor of $\sqrt{2}$.]
The corresponding predictions for the branching ratios of the CP-conjugate
decays are identical, and thus the associated CP asymmetries are
zero in this limit.
Measurements by the CLEO collaboration~\cite{Jessop:2000bv}
of the CP-averaged modes are consistent
with this prediction:
\begin{eqnarray}
{\cal B}({\bar B}^0 \to \rho^\mp K^\pm)  &=& (16.0
\stackrel{+7.6}{{\;\!}_{-6.4}} \pm 2.8) \cdot 10^{-6}  \\
{\cal B}({B}^- \to \rho^0 K^-) &=&  (8.4
\stackrel{+4.0}{{\;\!}_{-3.4}} \pm 1.8) \cdot 10^{-6} \;. \non
\end{eqnarray}
The more extensive data for the $B \to \pi K$
modes~\cite{CLEO00,BaBar01,Belle01} are summarized in
Table~\ref{tab:data}.
In this case the simple 2:1:2:1 pattern is
only roughly realized.
No significant CP asymmetries have been observed thus far.

Before proceeding to detailed estimates, we discuss the
contributions to $B\to \pi K$ and $B\to \rho K$ decays
in terms of a parametrization based on the Wick contractions in the
matrix elements of the operators
of the effective Hamiltonian~\cite{Buras:2000ra}.
The individual parameters that
appear are manifestly scale and scheme-independent, so that
the deficiencies of analyses based on the
factorization approximation~\cite{Buras:1999us} are avoided.
We first assume, as in Ref.~\cite{Buras:2000ra},
that only the valence degrees of freedom
of the mesons participate in the decay.
The various $B\to\pi K$ amplitudes are then parameterized as
\begin{eqnarray}
&&{\cal A}(B^0 \to K^+\pi^-) =
V_{us}V_{ub}^\ast(E_1(s,u,u;B^0,K^+,\pi^-)
- P_1^{\rm GIM}(s,u;B^0,K^+,\pi^-)) \nonumber\\ &&
- V_{ts}V_{tb}^\ast P_1(s,u;B^0,K^+,\pi^-)
\label{vampimkp}
\\
&&{\cal A}(B^+ \to K^+\pi^0) =
\frac{V_{us}V_{ub}^\ast}{\sqrt{2}}(E_1(s,u,u;B^+,K^+,\pi^0)
+ E_2(u,u,s;B^+,\pi^0,K^+) \nonumber\\
&&- P_1^{\rm GIM}(s,u;B^+,K^+,\pi^0)
+ A_1(s,u,u;B^+,K^+,\pi^0)) \nonumber\\
&&- \frac{V_{ts}V_{tb}^\ast}{\sqrt{2}}P_1(s,u;B^+,K^+,\pi^0)
+ \Delta A(B^+\to K^+ \pi^0)
\label{vampi0kp}
\\
&&{\cal A}(B^+ \to K^0\pi^+) =
V_{us}V_{ub}^\ast(A_1(s,d,u;B^+,K^0,\pi^+) - P_1^{\rm GIM}(s,d;B^+,K^0,\pi^+))
\nonumber \\
&&- V_{ts}V_{tb}^\ast P_1(s,d;B^+,K^0,\pi^+)
\label{vampipk0}
\\
&&A(B^0 \to K^0 \pi^0) =
 \frac{V_{us}V_{ub}^\ast}{\sqrt{2}}(
E_2(u,u,s;B^0,\pi^0,K^0)
+ P_1^{\rm GIM}(s,d;B^0,K^0,\pi^0))
\nonumber \\
&& - \frac{V_{ts}V_{tb}^\ast}{\sqrt{2}}P_1(s,d;B^0,K^0,\pi^0)
+ \Delta A(B^0\to K^0 \pi^0) \;.
\label{vampi0k0}
\end{eqnarray}
We have used the notation of Ref.~\cite{Buras:2000ra} and the convention
 $\rho^-, \pi^- ={\bar u}d$ and
$\rho^0, \pi^0 = ({\bar u}u - {\bar d}d)/\sqrt{2}$.
The corresponding parametrization of the $B\to \rho K$ amplitudes
is obtained from the replacement $(\pi^-,\pi^0,\pi^+) \to
(\rho^-,\rho^0,\rho^+)$.
The label ``$E_i$'' refers to operators with  $W^-$ emission topologies;
``$A_i$'' refers to annihilation topologies, whereas
``$P_i$'' contains penguin topologies.
The term ``$P_1^{\rm GIM}$'' represents penguin contributions which
vanish in the $m_c=m_u$ limit.
The contribution labelled ``$\Delta A$'' vanishes in the limit
of isospin symmetry; electroweak penguin effects contribute to it,
as do isospin-violating contributions in
the matrix elements themselves~\cite{isocom}.

\begin{figure}
\begin{center}
\epsfig{file=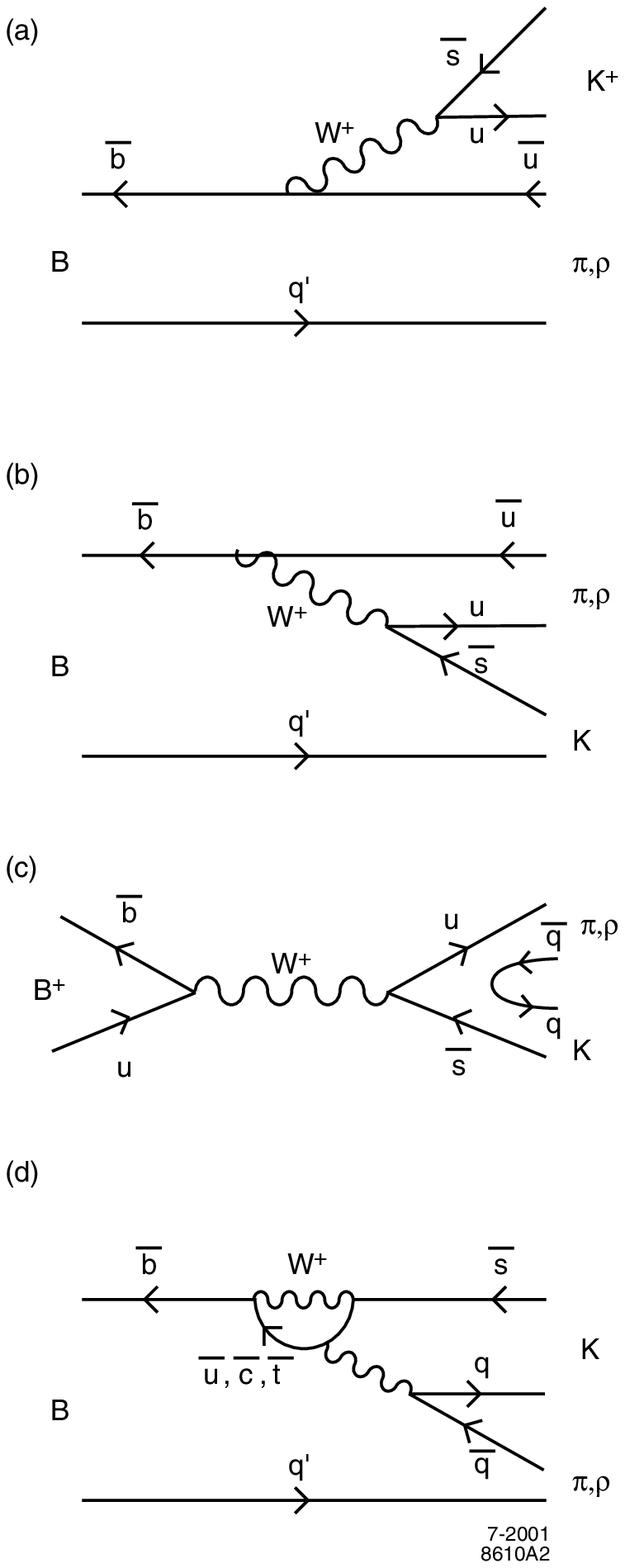, height = 20cm}
\end{center}
\caption{
Schematic illustrations of the full SM contributions
to the amplitudes parametrized in Eqs.~(\protect\ref{vampimkp}-
\protect\ref{vampi0k0}).
Diagram a) contributes to $E_1(s,u,u,B,K^+,\pi/\rho)$, diagram b)
contributes to $E_2(u,u,s,B,\pi/\rho,K)$,
and  diagram c) contributes to $A_1(s,q,u,B^+,K,\pi/\rho)$.
Diagram d) contributes to 
$P_1(s,q,B,K,\pi/\rho)$ and
$P_1^{GIM}(s,q,B,K,\pi/\rho)$.
}
\label{fig2}
\end{figure}

In the effective theory, the effects of the heavy
degrees of freedom, such as the $W^\pm, Z$ or $t$ quark are replaced by
effective coupling constants, the Wilson coefficients $C_i(\mu)$,
multiplying effective vertices $O_i(\mu)$.
Combinations of the products of $C_i(\mu)$ and $O_i(\mu)$ are
individually scale and scheme invariant.  Nevertheless,
the physics of the diagrams of the full SM
remains, and in Fig.~\ref{fig2} we illustrate the schematic diagrams
in the full theory which underlie the effective vertices and parameters.

We now enlarge 
our considerations to include
non-valence degrees of freedom in the meson wave functions.
The form of the parametrization itself does not change,
though additional terms arise from the decay
processes which do not appear in valence approximation.
Turning to Eq.(5) in Ref.~\cite{Buras:2000ra}, we see, adopting
their conventions [The numerical values of the $C_i(\mu)$
can depend on the explicit form of the operators~\cite{Buras:1992jm}, but this
impacts neither the identification of the effective parameters
nor their numerical values~\cite{Buras:2000ra}.], that the terms of form
\be
\frac{G_F}{\sqrt{2}} V_{cb}^\ast V_{cs}[C_1(\mu)Q_1^{scc}(\mu)
+ C_2(\mu)Q_2^{scc}(\mu)]
\ee
can contribute to $B\to \pi K$ and $B\to \rho K$ decay once
IC in the hadron light-cone wave functions is considered.
Such terms are Cabibbo-enhanced, and contain a Wilson coefficient
of order unity, so that
the phenomenological impact of these neglected terms
can be substantial.

\begin{figure}
\begin{center}
\epsfig{file=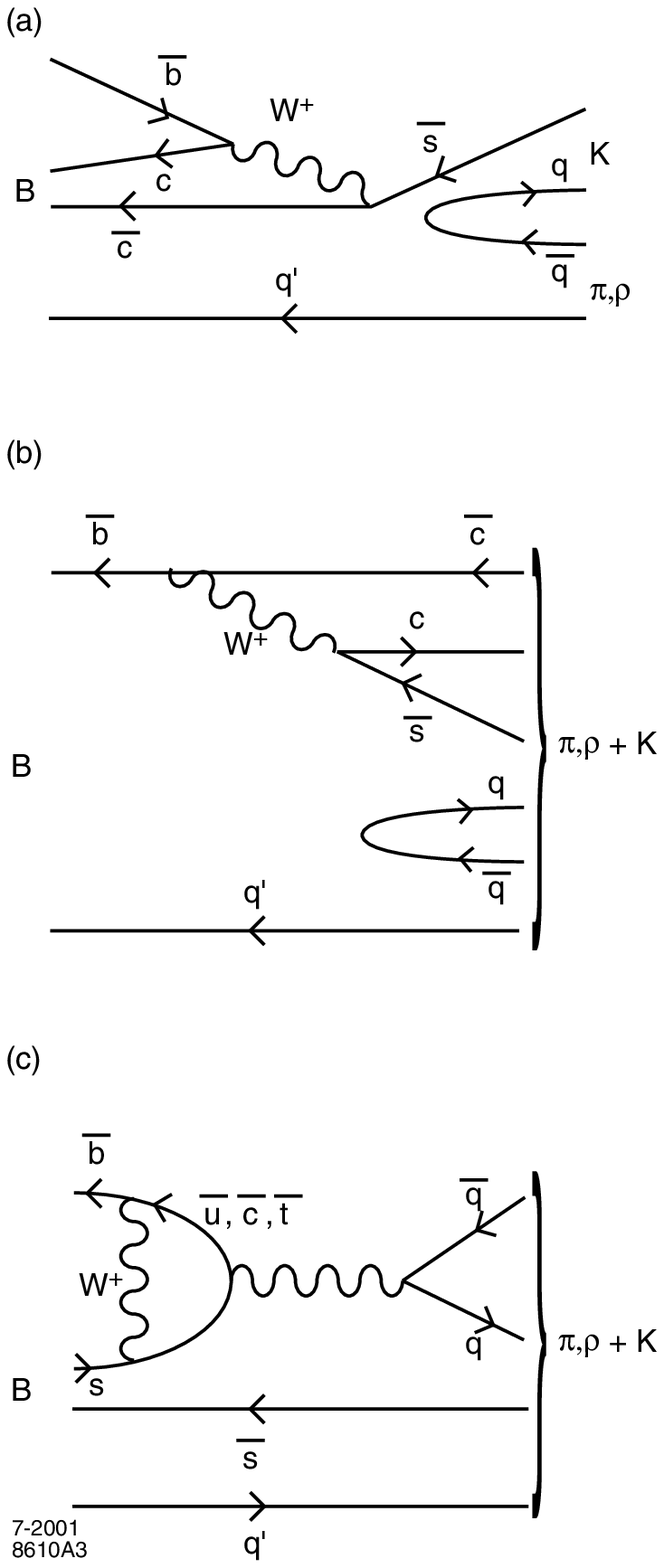, height = 20cm}
\end{center}
\caption{
Schematic illustrations of the full SM contributions
to $B\to K \,\pi/\rho$ decay as mediated by intrinsic charm
and strangeness in the hadron light-cone wave functions.
Diagrams a) and b) contribute to
$A_1^{IC}(s,q,B,K,\pi/\rho)$, whereas diagram c) modifies
the value of $P_1(s,q,B,K,\pi/\rho)$ and
$P_1^{GIM}(s,q,B,K,\pi/\rho)$ determined in the valence approximation.
}
\label{fig3}
\end{figure}

Let us define
\be
A_1^{\rm IC}(s,q;B,M_1,M_2) =
C_1 \langle M_1 M_2| O_1^{c} | B \rangle_{\rm IC}+
C_2 \langle M_1 M_2 | O_2^{c} | B \rangle_{\rm IC}\;,
\ee
where $q\in u,d$.
The contributions to $A_1^{\rm IC}(s,q;B,M_1,M_2)$
arising from intrinsic charm are
shown in a) and b) of Fig.~\ref{fig3} ---
we anticipate that contribution b),
which is driven by the IC component of the Fock states of the
light hadrons, is particularly significant for $\rho K$ final states.
Thus, the parameter
$A_1^{\rm IC}(s,q;B,M_1,M_2)$ will depend on
whether $\rho K$ or $\pi K$ final
states are considered because the intrinsic
charm content of the $\pi$ and $\rho$ are most likely different.

Fig.~\ref{fig3}c) illustrates
how intrinsic strangeness (IS) can modify the $P_1$ and $P_1^{\rm GIM}$
contributions --- the decay topology indicated is realized
from the ``$CP(c,s,u;B,M_1,M_2)$ (connected penguin)''
topology of Ref.~\cite{Buras:2000ra}
by pulling the strange quark line ``backwards'' into a $s\bar s$
pair. This pictorial description is not meant to trivialize
the IS contribution: the latter is irrevocably associated with the
bound state's structure, as it is entangled with the other
quarks of its Fock component by two or more gluon attachments.

The essential point of the parametrization of Ref.~\cite{Buras:2000ra}
is that the parameters given therein are both
scale and scheme independent. In particular,
the emission topologies, $E_i$, are scale and scheme-independent,
irrespective of any penguin contributions, since we can consider
transitions in which all four quark flavors are different: in such
processes penguin contributions simply do not occur, and the
$E_i$ represent the physical amplitudes of the channels in question.
Similarly, decay channels exist  for which only annihilation
topologies contribute; thus the annihilation contributions
associated with the $O_1(\mu)$ and $O_2(\mu)$ are themselves scale
and scheme independent. Here we consider an annihilation topology
driven by the IC
components of the hadrons' wave functions;
thus, this contribution, too, is separately scale and scheme independent
--- and the additive parameter $A_1^{\rm IC}(s,q;B,M_1,M_2)$
parameterizes its contribution. In the limit of isospin symmetry,
one value of $A_1^{\rm IC}(s,q;B,M_1,M_2)$ characterizes the
$\pi K$ final states and another characterizes those of
$\rho K$. We can now modify our earlier parametrization to include
the presence of IC:
\begin{eqnarray}
&&{\cal A}(B^0 \to K^+\pi^-) =
V_{us}V_{ub}^\ast(E_1(s,u,u;B^0,K^+,\pi^-)
- P_1^{\rm GIM}(s,u;B^0,K^+,\pi^-)) \nonumber\\ &&
- V_{ts}V_{tb}^\ast P_1(s,u;B^0,K^+,\pi^-)
+  V_{cs}V_{cb}^\ast A_1^{\rm IC}(s,u;B^0,K^+,\pi^-)
\label{ampimkp}
\\
&&{\cal A}(B^+ \to K^+\pi^0) =
\frac{V_{us}V_{ub}^\ast}{\sqrt{2}}(E_1(s,u,u;B^+,K^+,\pi^0)
+ E_2(u,u,s;B^+,\pi^0,K^+) \nonumber\\
&&- P_1^{\rm GIM}(s,u;B^+,K^+,\pi^0)
+ A_1(s,u,u;B^+,K^+,\pi^0)) \nonumber\\
&&- \frac{V_{ts}V_{tb}^\ast}{\sqrt{2}}P_1(s,u;B^+,K^+,\pi^0)
+  \frac{V_{cs}V_{cb}^\ast}{\sqrt{2}} A_1^{\rm IC}(s,u;B^+,K^+,\pi^0)
+ \Delta A(B^+\to K^+\pi^0)
\label{ampi0kp}
\\
&&{\cal A}(B^+ \to K^0\pi^+) =
V_{us}V_{ub}^\ast(A_1(s,d,u;B^+,K^0,\pi^+) - P_1^{\rm GIM}(s,d;B^+,K^0,\pi^+))
\nonumber \\
&&- V_{ts}V_{tb}^\ast P_1(s,d;B^+,K^0,\pi^+)
+  V_{cs}V_{cb}^\ast A_1^{\rm IC}(s,d;B^+,K^0,\pi^+)
\label{ampipk0}
\\
&&A(B^0 \to K^0 \pi^0) =
 \frac{V_{us}V_{ub}^\ast}{\sqrt{2}}(
E_2(u,u,s;B^0,\pi^0,K^0)
+ P_1^{\rm GIM}(s,d;B^0,K^0,\pi^0)
\nonumber \\
&& + \frac{V_{ts}V_{tb}^\ast}{\sqrt{2}}P_1(s,d;B^0,K^0,\pi^0)
-  \frac{V_{cs}V_{cb}^\ast}{\sqrt{2}} A_1^{\rm IC}(s,d;B^0,K^0,\pi^0)
+ \Delta A(B^0\to K^0\pi^0)\;,
\label{ampi0k0}
\end{eqnarray}
as earlier the $\rho K$ final states are realized by the
replacement $\pi \to \rho$. The structure of these relations
show us that under the assumption of the unitarity of the CKM matrix 
IC enters in a manner {\it identical} to that
of the penguin contributions $P_1$ and $P_1^{\rm GIM}$ ---
so that the contributions cannot be distinguished phenomenologically.
The numerical size of the penguin contributions has been
debated in the literature: in particular, 
the penguin contraction of the $O_2^{c}$ operator,
the ``charming'' penguin,
entering $P_1$ and $P_1^{\rm GIM}$, can be 
enhanced by non-perturbative 
effects~\cite{Colangelo:1989gi,Buras:1994pb,Ciuchini:1997hb,Ciuchini:1998rj,Ciuchini:2001gv,Isola:2001ar}.
Recently, moreover, Ciuchini et al. have argued that 
any phenomenological deficiencies of the ``QCD factorization''
approach,
without annihilation contributions, in describing the $B\to \pi K$
branching ratios can be rectified by introducing an additive
phenomenological contribution to $P_1$~\cite{Ciuchini:2001gv}.
It is evident that such a phenomenological treatment will
mimic the impact of IC! Thus we see that charming
penguins need not be penguins at all. Ciuchini et al. have
discussed that a variety of physical effects, such as annihilation
contributions, e.g., could be covered by the ``charming penguin''
aegis; we have shown that IC is another effect, non-perturbative
in nature, and potentially of substantial numerical size, which
contributes in an identical manner.

The recent phenomenological analysis of Ciuchini et al. is
of high interest~\cite{Ciuchini:2001gv},
though it is unsatisfying: it is evident
that annihilation terms, which they neglect,
are numerically important~\cite{lks,Beneke:2001ev}. Thus we
need a theoretical framework in which the annihilation
contributions, including those associated with IC, can be
estimated. To do this, we adopt the perturbative QCD
treatment of Li, Yu, and
collaborators~\cite{hnli,Li:1995zm,Chang:1997dw,Yeh:1997rq},
to which we now turn.

\section{Perturbative QCD Framework}

In the usual perturbative QCD treatment of exclusive processes,
the amplitude for a particular exclusive process is formed
by the convolution of the nonperturbative distribution amplitudes,
$\phi_H(x,Q)$, with the hard scattering amplitude $T_H$ computed
from the scattering of on-shell, collinear
quarks~\cite{Lepage:1979zb,Lepage:1980fj}. To wit, for $B\to M_1 M_2$,
we have
\begin{equation}
\label{classic}
{\cal M}(B \to M_1 M_2) = \int_0^1 dz \int_0^1 dy \int_0^1 dx
\phi_B(x,Q) T_H(x,y,z) \phi_{M_1}(y,Q) \phi_{M_2}(z,Q) \;,
\end{equation}
where, e.g.,
$\phi_{M_2}(z,Q)=\int_0^{Q} d^2 k_\perp \phi(x,k_\perp,\lambda_i)$.
This formula is suitable if the
distribution amplitudes vanish
sufficiently rapidly at the endpoints, and if $\alpha_s(\mu)$ is sufficiently
small for a perturbative treatment to be germane.
However, in the case of
some electromagnetic form factors~\cite{isgur},
as here in B decay~\cite{Szczepaniak:1990dt,Burdman:1991hg}, the
distribution amplitudes may not fall sufficiently quickly at the
endpoints to permit both criteria to be satisfied.
Eq.~(\ref{classic}) itself emerges from an expansion
of $T_H$ in powers of $k_\perp^2/Q^2$; the solution~\cite{Botts:1989kf,lipi}
is to reorganize the contributions in $k_\perp$, so that the
contributions to the hard scattering in the transverse configuration
space ($b$, conjugate to $k_\perp$) are no longer of point-like size.
The $b$ dependence of the reorganized distribution amplitudes,
the so-called ``Sudakov exponent,'' suppresses the large $b$ region,
so that the resulting integrals are convergent and $\alpha_s(\mu)$
is more or less consistently small. As $Q^2$ increases, the
Sudakov mechanism becomes more effective at screening the 
large $b$ 
region~\cite{lipi}. The soft portion of the integrand can also be
regulated by adopting a ``frozen'' running coupling for sufficiently
low scales~\cite{Ji:1987uh} or by incorporating transverse structure
in the light-cone wave functions~\cite{lech}.
The procedure of Ref.~\cite{hnli,Li:1995zm} is appealing in its
simplicity as a single parameter, $\Lambda_{QCD}$,
suffices to regulate the nonperturbative dynamics.

\begin{figure}
\begin{center}
\epsfig{file=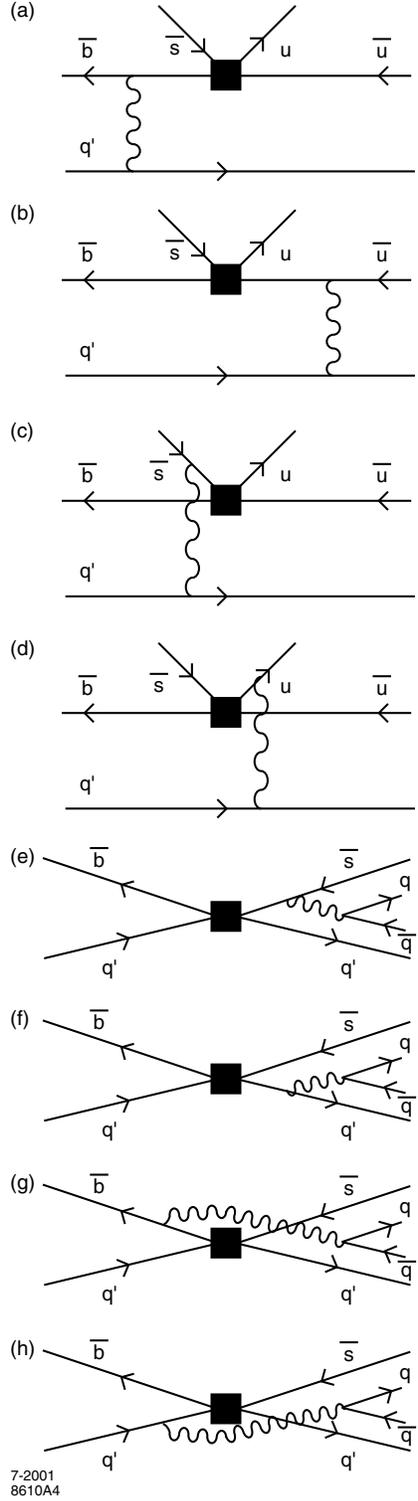, height = 20cm}
\end{center}
\caption{
Hard scattering diagrams in the valence approximation to
$B\to K (\pi/\rho)$ decay.
}
\label{fig4}
\end{figure}

In the approach of Refs.~\cite{hnli,Li:1995zm,Chang:1997dw,Yeh:1997rq}
the transition amplitude for $B\to M_1 M_2$ decay is estimated via
a three-scale factorization
formula, typified by the convolution
\begin{equation}
C(t)\otimes H(t) \otimes \phi(x,b) \otimes \exp\left[
-s(P,b) - 2 \int_{1/b}^t \frac{d\bar{\mu}}{\bar{\mu}}
\gamma_q[\alpha_s(\bar{\mu})] \right]\,,
\end{equation}
where $\gamma_q=-\alpha_s/\pi$ is the quark anomalous dimension
in axial gauge and $t\sim {\cal O}(m_B)$. The
Wilson coefficient $C(t)$ reflects the RG evolution of a term of
the effective Hamiltonian,
such as given in Eq.~(\ref{heff}), from the W mass scale, $M_W$,
to the hard scale $t$. The light-cone wave function $\phi(x,b)$,
on the other hand,
parameterizes the nonperturbative dynamics manifest at scales
below $1/b\sim {\cal O}(\Lambda_{QCD})$. The exponential
factor organizes the large logarithms which occur when the
system is evolved from the scale $t$ to that of $1/b$; this
contribution includes the
Sudakov form factor, with exponent $s(P,b)$ and $P$ the
dominant light-cone component of a meson momentum, and
suppresses the contribution of the large $b$ region.
The remaining part, the hard scattering amplitude $H(t)$,
is channel-dependent and can be calculated
perturbatively using the operators $O_i$ of the effective
Hamiltonian. The diagrams which comprise $H(t)$
for $B\to \pi K$ decay in ${\cal O}(\alpha_s)$
are shown in Fig.~\ref{fig4} --- we consider specifically
$B\to \pi K$ decay in this section
as this system has recently been treated by
Li, Keum, and Sanda in the perturbative QCD approach~\cite{lks}.
These are the only diagrams in ${\cal O}(\alpha_s)$;
contributions without
a hard-gluon exchange between the spectator and other quarks are
suppressed by wave functions and do not
contribute~\cite{Szczepaniak:1990dt,Lu:2001em}.

Let us consider the leading-order
contributions to $H(t)$ in  $B\to \pi K$ decay.
In diagrams a) and b) the K meson
is produced in a color singlet state, so that the lower
half of the process ``factorizes'' from the meson emission
and represents a computation of the $B\ra \pi$ form factor.
In diagrams c) and d) the weak process produces the mesons
in a color octet state; the exchanged, hard gluon ensures that the
outgoing mesons emerge in color-singlet states.
Finally, diagrams e)-h) describe annihilation processes; the gluon emission
produces a $q\bar q$ pair, so that two color-singlet mesons can
be produced. The limitations of the usual factorization
formula, Eq.~(\ref{classic}), become apparent upon the evaluation
of the diagram in
Fig.~\ref{fig4}a)~\cite{Szczepaniak:1990dt,Burdman:1991hg,Akhoury:1994uw,hnli}
for the $B\to \pi$ form factor. If the
transverse momentum dependence of the quarks is neglected,
the heavy $b$ quark generates
a singularity as $x_3$, the longitudinal momentum fraction of
the spectator quark in the $\pi$-meson, goes to zero~\cite{kinem}. However,
this singularity no longer occurs if the $k_\perp$ dependence
in the heavy quark propagator
is retained; thus
the $B\to \pi$ form factor is regarded as perturbatively calculable,
once proper resummation techniques are applied~\cite{hnli}.
In the approach of Beneke {\it et al.}~\cite{Beneke:1999br}, however,
the $B\to \pi$ form factor is treated as
non-perturbative input. One consequence of this is that
the perturbative corrections in the two schemes are
organized differently:
in Ref.~\cite{Beneke:1999br}, hard-scattering contributions in
${\cal O}(\alpha_s)$ are retained
which would be regarded as non-leading order in the approach
of Ref.~\cite{hnli}. However, infrared enhancements also plague
the treatment of annihilation contributions in this
approach~\cite{Beneke:2001ev};
thus the treatment of Ref.~\cite{hnli} is more systematic
in that the endpoint singularities
contributions from the contributions of all decay
topologies are regulated in a consistent way.
For further comparison of the two approaches, see
Refs.~\cite{Keum:2001ms,Beneke:2001ev,Brodsky:2001jw}.

Adopting the notation and conventions of Ref.~\cite{lks} we
find that the parameters of Eqs.~(\ref{vampimkp}-\ref{vampi0k0})
are given by~\cite{details2}
\begin{eqnarray}
E_1(s,u,u;B,K,\pi) &=& -f_K F_e - M_e \,,\nonumber \\
E_2(u,u,s;B,\pi,K) &=& -f_\pi F_{eK} - M_{eK} \,, \nonumber \\
A_1(s,u,u;B,K,\pi) &=& -f_B F_a - M_a \,,
\label{mapping}\\
P_1(s,q,B,K,\pi) &=& -f_K F_e^P - M_e^P - f_B F_{a}^P - M_{a}^P \,,\nonumber \\
P_1^{\rm GIM}(s,q,B,K,\pi) &=& 0 \,, \nonumber \\
\Delta A(B\to \pi K) &=&  V_{ts}V_{tb}^\ast
(f_\pi F_{eK}^P + M_{eK}^P) \,, \nonumber
\end{eqnarray}
where ``factorizable'' and ``non-factorizable'' contributions
are denoted by $F$ and $M$, respectively. The subscripts $e$ and $a$ 
refers to emission and annihilation topologies, respectively, and the
$P$ superscript reflects the presence of penguin operators 
in the hard-scattering amplitude.
Diagrams a) and b) in Fig.~\ref{fig4} give rise
to the contribution $f_K F_e^{(P)}$, whereas
Figs.~\ref{fig4} c) and d) give rise to $M_e^{(P)}$. ``Factorizable''
in this context means that the production of one meson is independent
of that of the other, so that as illustrated in diagrams a) and b), e.g.,
the emission of the $K$-meson decouples
from the dynamics of the $B\to \pi$ form factor.
Switching the $\bar{s}$ and $\bar{u}$ labels in diagrams
a)-d) of Fig.~\ref{fig4},
one finds that the new diagrams
a) and b) give rise
to the contribution $f_\pi F_{eK}^{(P)}$, whereas
the new diagrams c) and d) give rise to $M_{eK}^{(P)}$.
Note, moreover, diagrams
e) and f) in Fig.~\ref{fig4} give rise to
$f_B F_A$ whereas g) and h) give rise to $M_A$. Note that
$q\in (u,d)$; the contributions
$F_e^P$, $M_e^P$, $F_a^P$, and $M_a^P$
subsume quark-charge-dependent pieces. Finally,
electroweak penguin contributions give rise to the
terms denoted by $\Delta A(B\to \pi K)$.

The explicit expressions for the various $F$
and $M$ are detailed in Ref.~\cite{lks} and references therein.
The best fit of their resulting branching ratios to the data
of Ref.~\cite{CLEO00} suggests that $\gamma=\phi_3\approx 90^\circ$.
Combining the CLEO~\cite{CLEO00} results with the
more recent measurements at BaBar~\cite{BaBar01} and Belle~\cite{Belle01},
noting Table \ref{tab:data}, suggest that $\gamma$ can be smaller
than $90^\circ$, a result in closer accord to the value of
$\gamma$ determined from global
fits of the CKM matrix in the
SM~\cite{Hocker:2001xe,Ciuchini:2000de}.
Significant CP asymmetries in the $B^\pm \to K^\pm \pi^0$
and $B^0 (\bar{B^0}) \to K^\pm \pi^\mp$ modes are also predicted.
The branching ratios in these modes are 
sensitive to $\gamma$. However, we shall show
that the IC contribution may impact this picture in a significant way.

Significant CP asymmetries in the $B^\pm \to K^\pm \pi^0$ 
and $B^0 (\bar{B^0}) \to K^\pm \pi^\mp$ decays
are predicted in Ref.~\cite{Keum:2001ph} since the factorizable
annihilation diagrams  generate significant strong phases through
imaginary parts in the hard scattering amplitudes.  
Li, Keum, and Sanda argue that the Sudakov suppression 
mechanism required to regulate the
endpoint regions of the amplitudes is reliable since
some 90\% of the contributions
to the $F_{e6}^P$ form factor comes from the region where
$\alpha_s/\pi < 0.3$~\cite{Keum:2001ph,lks}. 
This is an encouraging, if weak, self-consistency
check, though it is crucial to recognize that this remark pertains 
specifically to the
value of $\alpha_s$ associated with the gluon exchanged
in the computation of the hard scattering kernel. 
The function $\alpha_s(\bar\mu)$ also appears 
parametrically in the expression for the Sudakov exponents; here
$\alpha_s(\bar\mu)$ can be evaluated at the softest scale, namely
$\alpha_s(1/b)$ with $b\sim 1/\Lambda_{\rm QCD}^{(4)}$. 
The sensitivity of the numerical results to soft physics can thus
be codified in terms of the sensitivity of the results to the
cut-off in the integration over the transverse coordinates. 
In specific, if we change the cut-off from 
$1/\Lambda_{\rm QCD}^{(4)}$ to  $0.45/\Lambda_{\rm QCD}^{(4)}$, so
that $\alpha_s(\bar\mu)/\pi \le 0.30$ throughout the entire
integral, we find that the $F_e^6$ form factor changes by
a factor of three. The numerical results depend 
sensitively on the manner
in which the endpoint region is regulated. Hopefully the
inclusion of ``threshold resummation'' will help mitigate this
sensitivity~\cite{Li:2001ay}. 
It should also be emphasized that a more realistic 
regulator might comprise of the Sudakov suppression 
mechanism with a systematic 
scale setting procedure in concert with a ``frozen''
value of $\alpha_s$ at very low scales. In this manner, the
sensitivity of the numerical results to the endpoint regions
can be reduced, but we postpone this investigation
to a later publication. 

Let us now proceed to investigate the consequences of IC in
$B\to \pi K$ decay. The hard scattering contribution to
$B\to \pi K$ decay mediated by IC in the B-meson is shown in
Fig.~\ref{fig5}; the hard gluon must emerge from the spectator
valence quark to generate a non-trivial contribution.
 This diagram gives rise to the parameter
$A_1^{\rm IC}(s,q,B,K,\pi)$ in Eqs.~(\ref{ampimkp}-\ref{ampi0k0}).
It is worth noting that IC in the $K$ or the $\pi$ mesons could also
generate a contribution to $A_1^{\rm IC}(s,q,B,K,\pi)$, as illustrated
schematically in Fig.~\ref{fig3}b). We ignore this contribution,
as we believe that IC in the B-meson to be of greater
phenomenological significance. In order to estimate the contribution
associated with Fig.~\ref{fig5}, we must consider the structure
of the B meson in the presence of IC. It is to this issue we now turn.

\subsection{IC in the B Meson}

The B-meson light-cone wave function $\ket{\Psi_B}$ as an eigenstate
of its light-cone Hamiltonian $H_{lc}$
satisfies the equation
$(M_B^2 - H_{lc})\ket{\Psi_B} =0$.
Expanding $\ket{\Psi_B}$
in terms of its Fock state decomposition
yields an infinite set of integral equations for the
Fock components $\psi_{n}(x_i,\vec{k_\perp}_i,\lambda_i)$,
so that
\begin{equation}
\psi_{n}(x_i,\vec{k_\perp}_i,\lambda_i)
= \frac{1}{M_B^2 - \sum_i m_{\perp i}^2/x_i}
({\mathbf V_{lc}} \Psi_B)_n \;,
\label{bseqn}
\end{equation}
where ${\mathbf V_{lc}}$ is the potential matrix in the Fock-state
basis. The sum is over the partons in the Fock component $n$,
namely $\sum_i k_i^-$, and
$m^2_{\perp i}\equiv \vec{k}_{\perp i}^2 + m_i^2$.
The kinematics dictate $\sum_i x_i =1$.
Equation (\ref{bseqn}) suggests that
minimizing the value of $\sum_i m_{\perp i}^2/x_i$ enhances
the likelihood of a particular Fock-state component.
This is equivalent to the statement that the kinematics
of the particles in each Fock state tend to minimize the total
light-cone energy. Thus the most favored
configurations have zero relative rapidity and
$x_i \propto m_{\perp i}.$

We enumerate
the lowest-lying $\ket{b c \bar{c} \bar{q}}$ states consistent
with the $0^-$ B-meson state in Table \ref{tab:IC}, and
we assess their relative likelihood, that is, their relative
light-cone energy, by estimating the
total invariant
mass of each configuration of specific spin by summing the mass of
each individual
$(q_1\bar{q_2})_{J^P}$ state.  [For simplicity, we neglect the mass shift
associated with the relative motion of the $q_1 \bar{q_2}$ states.] To
effect this estimate,  we use the empirically determined
masses~\cite{pdg2000}, supplemented by model
calculations~\cite{Ebert:1998nk,Godfrey:1985xj} for those states which
have not yet been observed~\cite{details}.  The net result is that
the enumerated states listed in Table \ref{tab:IC} are of comparable
likelihood.

The pattern of estimated 
masses can be understood by noting that
in the heavy quark limit, the spin of the heavy quark decouples, and
the ground state is a degenerate
doublet of $(0^-,1^-)$ states~\cite{Isgur:1991wq}. There are two excited
P-wave doublets, comprised of $(0^+,1^+)$
and $(1^+,2^+)$ states;
an estimate of spin-orbit effects suggests that the $(1^+,2^+)$ doublet is
more massive~\cite{DeRujula:1976kk}.
More recent investigations have suggested,
however, that the sign of the spin-orbit
force in heavy-light systems is inverted relative to atomic
expectations~\cite{Isgur:1998kr}, so that
the $(0^+,1^+)$ doublet is pushed to
higher energy~\cite{Isgur:1998kr,Ebert:1998nk}; nevertheless,
the configurations of
two $1^+$ mesons in a relative $l=1$ state can be discounted, as they
are more massive and thus less likely than the states
enumerated in Table \ref{tab:IC}.
The combination of $0^-$ and $2^+$ mesons in a relative $l=2$ state
can be similarly neglected, as the masses of the mesons themselves
are comparable or slightly larger
than those of the $0^-$ and $0^+$ configurations --- and the
angular momentum penalty is two units.
All the IC configurations involve at least one  $l=1$ configuration,
either within or between the meson states. This
implies that the numerator of the  light-cone wavefunctions 
can involve terms which are 
linear in the relative transverse momenta
$k_{\perp i}$~\cite{Brodsky:2001ii}.

The recent observation of a significant branching ratio of $B$'s decaying
into the scalar $P$-wave state of charmonium~\cite{Abe:2001pf}
\begin{equation}
{{\cal B}(B^+ \to K^+ \chi_0) \over {\cal B}(B^+ \to K^+ J/\psi)}
= 0.77 \pm 0.11 \pm 0.22
\label{bellechi}
\end{equation}
is surprising from the standpoint of naive factorization since
the $V-A$ structure of the weak vertex forbids the decay to
$\chi_0$ and $\chi_2$ states~\cite{Beneke:1999ks,Diehl:2001xe}.
However, QCD radiative corrections
to the inclusive decay rate are 
sizeable~\cite{Beneke:1999ks}, so that 
once these corrections
are included, the decay rates into the $\chi_{0,2} K$ and $J/\psi K$
channels may not be dissimilar~\cite{Diehl:2001xe}.

On the other hand, the large observed decay rate for $B^+ \to K^+ \chi_0$
could be evidence for a transition involving intrinsic charm of the $B$
since the required orbital angular momentum is naturally present.  The
first four configurations in Table \ref{tab:IC} are suggestive of a
mechanism in which the $b\bar{q}$ annihilate to a K meson; the remaining
charmonium  state of either $1^-$, $0^-$, or $0^+$ character is freed
from the B-meson wave function.
Helicity arguments favor
annihilation in the
$1^-\,(b\bar{q})$ state, though
the suppression associated with annihilation in $0^-$ state
may not be severe~\cite{lks}.
Nevertheless, this mechanism populates the
$\eta_c K$, $J/\psi K$, and $\chi_0 K$ final states, though
the decay to the
$\chi_2$ state, however, is disfavored, offering a
signal which can distinguish intrinsic charm from other
mechanisms. Note that the weak annihilation of the $b\bar{q}$
quarks must be mediated by penguin operators
to avoid suppression by CKM factors.

The IC mechanism can occur in inclusive decays as well.
The likelihood of materializing IC Fock states is
enhanced at the value of $x_F$ which corresponds
to the typical $x$ of
the IC configuration in the parent hadron~\cite{Brodsky:1992dj};
in B-decay, this means that
$x_{c{\bar c}} \sim \langle x_c \rangle
+ \langle x_{\bar c} \rangle\approx 0.44$~\cite{Chang:2001iy}.
The kinematics, in turn, determine the invariant mass of $X_s$
for which this value of $x_{c\bar c}$ can be realized, namely
$M^2_B =
(M^2_{X_s}+ k^2_\perp)/(1-x_{c \bar c}) +
(M^2_{c \bar c} + k^2_\perp )/ x_{c \bar c}$, so that
the IC
mechanism is particularly favored in this region
of $X_s$ invariant mass.
Note that the decay
$B\to \chi_2 X_s$ continues to be disfavored. We can compare this
scenario with the
NRQCD prediction which emerges in leading order in the
relative velocity of the $c\bar c$ pair;
in this alternative picture, the B decays to the P-wave charmonia can
be mediated by a decay to a color-octet charmonium state, which evolves
non-perturbatively to a color-singlet $\chi_J$
state~\cite{Bodwin:1992qr,Beneke:1999ks}.
This mechanism, in concert with the color-singlet mechanism,
computed in NLO, suggests that
${\cal B}(B\to \chi_2 X_s)/
{\cal B}(B\to \chi_0 X_s) \sim 5$~\cite{Beneke:1999ks}.
Thus a comparison of the $B\to \chi_{0,2} X_s$ branching ratios
as a function of $X_s$
should serve to test and distinguish the IC and color-octet mechanisms.

Returning to Table \ref{tab:IC} we see that the $\bar D$ and the
$1^-$ and $0^+ \,\bar{D}^*$ states can also be liberated from the B-meson
via annihilation of the ${\bar b}c$ quarks.
Indeed, this mechanism
is illustrated for decay into the ${\bar D}^*$ $(1^-)$ state in Fig.2e) of
Ref.~\cite{Chang:2001iy} --- the annihilation can occur without
CKM suppression into $\bar{d}u$ and $\bar{s}c$ quarks, or into
$l^+ \nu_l$. Moreover, the annihilation in this case can
occur with a Wilson coefficient of order unity, so that the mechanism
is numerically more significant in this case than in decays
to charmonium final states.
The helicity-favored process can yield either a $\bar D$ or ${\bar D}^*$
meson. IC has been argued to impact the $B\to {\bar D}^*$ form factor
near the zero recoil region, and thus the extraction of
$|V_{cb}|$~\cite{Chang:2001iy}. The error in $|V_{cb}|$
is of the order of 5\%~\cite{pdg2000}; yet the
presence of IC could become an important consideration
as the form-factor studies gain
in precision. The
materialization of IC Fock states
is more prominent at
larger values of $x_F$; this argues for the materialization
of IC in this case at small recoil. Potentially the presence of
IC could impact the extrapolation of the empirically measured
$B\to {\bar D}^*$ form factor to zero recoil.
Interestingly, however, theoretical constraints exist on 
the slope and curvature of the $B\to {\bar D}^{(*)}$
form factors~\cite{Boyd:1997kz,Caprini:1998mu}, 
and thus a comparison with high-precision experimental data
may serve to identify the contribution from IC.
It is worth noting that
the decay to ${\bar D}_2^*$ is disfavored in the IC mechanism; comparing
branching ratios to ${\bar D}_2^* h$ and ${\bar D}_0^* h$ final states
could also be helpful in illuminating the mechanisms involved.
These notions
extend to the inclusive case as well, in a manner analogous to
our earlier discussion.

Let us now proceed to specific numerical estimates of the IC
mechanism in $B\to \pi K$.

\subsection{Numerical Estimates}

\begin{figure}
\begin{center}
\epsfig{file=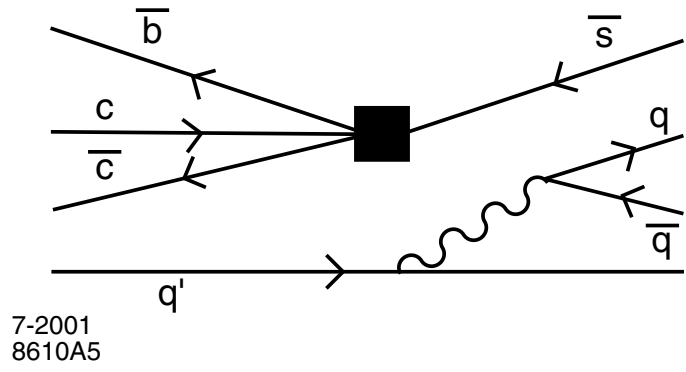, width = 9cm}
\end{center}
\caption{
Hard scattering diagram mediated by intrinsic charm
in the B-meson wave function.
}
\label{fig5}
\end{figure}

In constructing a
first estimate of the impact
of IC in $B\to \pi K$ decay we consider exclusively
the role played by IC in the B-meson since we expect it to have
larger phenomenological impact than the IC of the light
mesons. The hard
scattering diagram for $B\to \pi K$ decay in ${\cal O}(\alpha_s)$
is shown in Fig.~\ref{fig5}. The color structure of the
$O_2^{c}$ operator which
causes the weak decay also allows
diagrams where the   $g^* \to q\bar q$  originates
from the $\bar s$ or $\bar c$ quark line.
The light-cone-energy minimization
argument suggests that $\langle x_q \rangle$ is pushed to
very low x, as in Fig. 3b) of Ref.~\cite{Chang:2001iy}, so
that the spectator quark must suffer a hard interaction
to emerge in the pion final state with an appreciable
likelihood. Thus these additional contributions
are suppressed by the hadron wave functions, and we need not
consider them further.

Although IC in the B meson can be in any of
several low-lying configurations, as illustrated in Table \ref{tab:IC},
we assume that our estimated lowest energy
state, namely the $(b\bar c)_{1^-}(c\bar q)_{0^-}$ state,
reflects the dominant arrangement of IC in the B meson wave function.
Examining Fig.~\ref{fig5} we see that there are no hard gluon interactions
across the weak vertex, so that the computation of the hard scattering
amplitude factorizes. One portion of the weak vertex
mediates the annihilation of the $1^-$ ${\bar b} c$ quarks, and the
other describes the amplitude for the $0^-$ $(c\bar q)$ state
to emerge with the parton content of the $\pi K$ final state,
namely ${\bar s} q' q {\bar q}$.
It should be realized, however, that the
final contribution is non-factorizable
since the two pieces of the hard scattering amplitude are tied
together by the integration over the
coordinates of the light-cone wave function
of the $(b\bar c)_{1^-}(c\bar q)_{0^-}$ state. We know little
about the form of this wave function, although the kinematics
of the IC configuration suggests that it falls more steeply in
the endpoint region than the wave function of the valence component.
Consequently, reorganizing the perturbative contribution
into integrals over the impact parameters may not be needed.
Nevertheless, to effect an explicit, albeit simple, estimate
we turn to the form factors computed by Ref.~\cite{lks}, as
the amplitude for the
$0^-$ $(c\bar q)$ state to emerge as
${\bar s} q' q {\bar q}$ cosmetically resembles Fig.~\ref{fig4}f).
We thus estimate that
\begin{equation}
A_1^{\rm IC}(s,q,B,K,\pi) \sim
f_{B_c^\ast} F_a^{P} \frac{a_1(m_b/2)}{a_6(m_b/2)}B \;,
\label{ICestimate}
\end{equation}
where $f_{B_c^\ast}\sim 0.317$ GeV\cite{Colangelo:1993cx} reflects
the annihilation of the $1^-$ $\bar b c$ quarks, and the remaining
factors reflect an estimate of the lower half of the diagram
in Fig.~\ref{fig5}. The $F_a^P$ form factor is dominated by the
$F_{a6}^P$, engendered by the operators associated with
$a_6(t)$, for which
the contributions from Figs.~\ref{fig4} e) and f)
sum, rather than approximately cancel.
We note that ${a_1(m_b/2)}/{a_6(m_b/2)}\sim -20$. The
parameter $B$ reflects the probability amplitude to find the
B meson in an IC configuration, as well as an adjustment for the
$\sim$ 50\% penguin enhancement reported in Ref.~\cite{Keum:2001ph}, so
that $B\sim 2(0.02)/3$. We take this as a rough estimate of the possible
impact of IC in $B\to \pi K$ decay.

\begin{figure}
\begin{center}
\epsfig{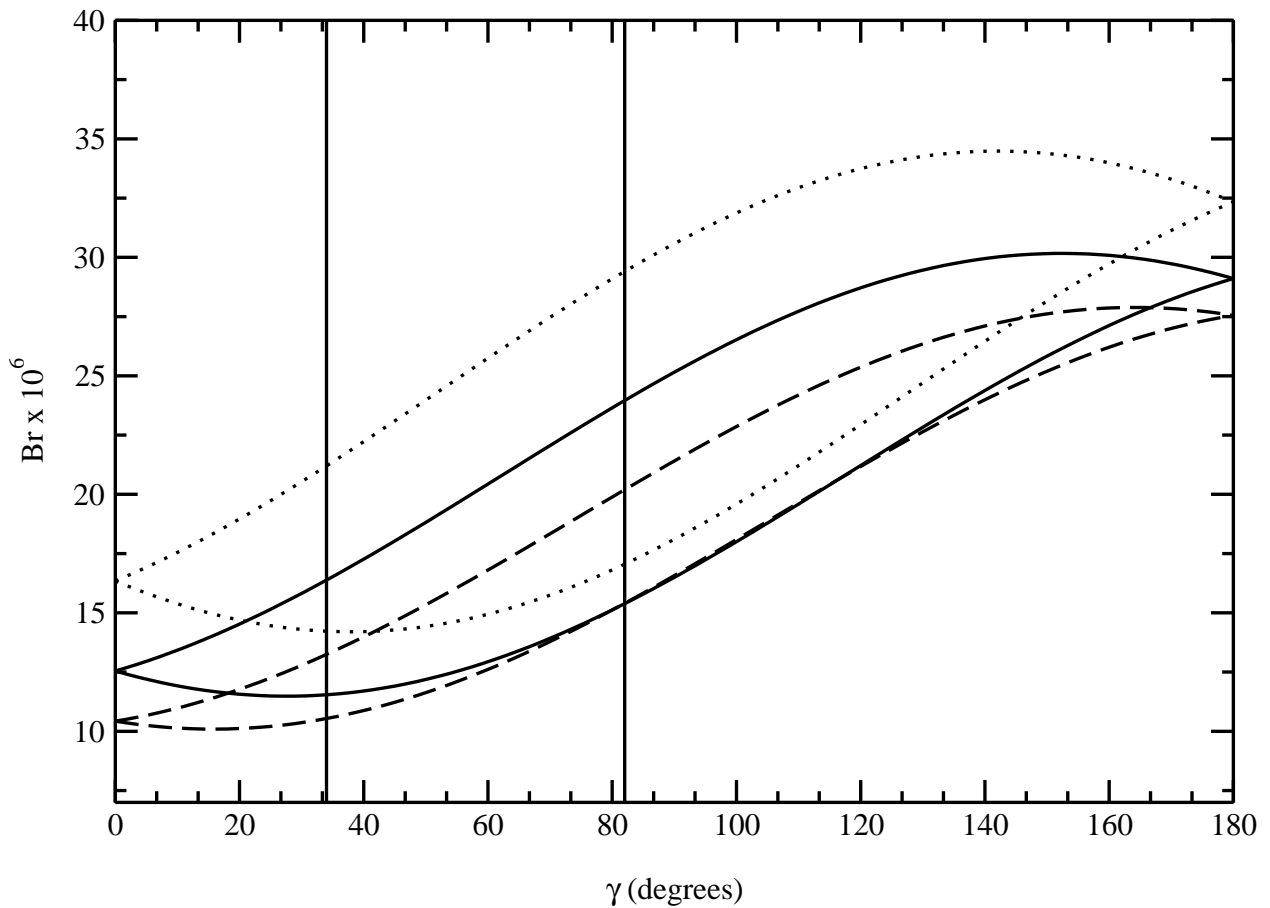}
\end{center}
\caption{
The impact of IC in the B meson on the
$B\to K^\pm\pi^\mp$ branching ratios 
as a function of $\gamma$. The {\it upper} curve for each type
of line corresponds to $B^0\to K^+\pi^-$ decay, whereas
the {\it lower} curve for each type of line corresponds to
${\bar B}^0\to K^- \pi^+$ decay. The solid line depicts the
results of Ref.~\protect\cite{lks}, realized from their Table I.
The dashed line
is the result once the IC contribution,
as per  Eqs.~({\protect\ref{ampimkp}},{\protect\ref{ICestimate}}),
is subtracted, and the dotted line
is the result once the IC contribution,
as per  Eqs.~({\protect\ref{ampimkp}},{\protect\ref{ICestimate}}),
is added. The vertical solid lines enclose the $\ge 5\%$ C.L.
fits to $\gamma$ in the SM,
$34^\circ \ge \gamma \ge 82^\circ$, of Ref.~\protect\cite{Hocker:2001xe}.
}
\label{fig6}
\end{figure}

In evaluating the role of IC in $B\to \pi K$ decays, let us
first consider what we might expect in its absence.
We can relate the parameters of
Eqs.~(\ref{ampimkp}-\ref{ampi0k0}) to the older, diagrammatic
analyses of Ref.~\cite{Gronau:1994bn}. To wit, the parameter
$E_1$ contains the factorizable, ``color-allowed'' tree
contribution, whereas $E_2$ contains the factorizable, ``color-suppressed''
tree contribution --- this is made manifest
in the mapping to the amplitudes of
the perturbative QCD approach in Eq.~(\ref{mapping}). Typically
the factorizable, ``color-suppressed'' tree and annihilation
--- specifically the $f_B F_a$ term in
Eq.~(\ref{mapping})) --- contributions are thought to be
smaller than the factorizable, ``color-allowed'' tree
contribution~\cite{Gronau:1994bn}. The explicit numerical estimates
of Ref.~\cite{lks,Beneke:2001ev} support this in that the partial
rate asymmetries in the modes with charged kaons, which contain
the $E_1$ parameter, are much larger. For the purpose of our
discussion, let us define the partial rate asymmetry in $K^\pm\pi^\mp$
decay as
\begin{equation}
A_{\rm CP}(K^+ \pi^-) \equiv \frac{
{\cal B}(B^+ \to K^+ \pi^-)
- {\cal B}(B^- \to K^- \pi^+)}{{\cal B}(B^+ \to K^+ \pi^-)
+ {\cal B}(B^- \to K^- \pi^+)}\;,
\end{equation}
where the extension to other modes is clear. We have argued (and
explicit calculations suggest) that
$A_{\rm CP}(K^+ \pi^-)$ and $A_{\rm CP}(K^+ \pi^0)$
should be larger than $A_{\rm CP}(K^0 \pi^+)$ and $A_{\rm CP}(K^0
\pi^0)$.  Sizeable asymmetries are generated when the $E_1$ parameter is
included; otherwise they are small.
The IC contribution, 
$A_1^{\rm IC}$, can be significant
relative to the penguin contribution, but it is small compared
to $E_1$, so that $A_{\rm CP}(K^0 \pi^+)$ and $A_{\rm CP}(K^0 \pi^0)$
will continue to be small once the effects of IC are included, although
they will be modified.
Consequently we focus on the role of IC in $B\to K^\pm \pi^\mp$
decay.
Using Eqs.~(\ref{ampimkp},\ref{mapping}) and the numerical
results of Table I in Ref.~\cite{lks} we find that
\begin{eqnarray}
  A_1^{\rm IC} &=  2.5\cdot 10^{-4}  + i\, 2.2\cdot 10^-3 \nonumber \\
  E_1          &= -0.13  -i\, 4.1\cdot 10^{-3} \\
  P_1          &=  9.3\cdot 10^{-3} - i\,4.5\cdot 10^{-3} \;, \nonumber
\end{eqnarray}
so that $|A_1^{\rm IC}|/|P_1| \sim 20\%$. The
resulting branching ratios, ${\cal B}(B^0\to K^+\pi^-)$ and its
CP-conjugate,
are plotted as a function of $\gamma$ in Fig.~\ref{fig6}.
The inclusion of IC modifies the expected branching ratios, but
its impact on the expected partial rate asymmetries is striking.
IC modifies the penguin contribution,
parameterized by $P_1$, substantially; in
realizing $A_{\rm CP}$ this effect is
amplified by the comparatively
large value of $E_1$, generating the substantial variations seen.
For similar reasons, one can also expect marked effects in
$A_{\rm CP}(K^+\pi^0)$.
The combinations of Wilson coefficients $a_1$ and $a_6$ differ
in sign, as do the ${\cal O}(\lambda^2)$ pieces of
$V_{cs} V_{cb}^\ast$
and $V_{ts} V_{tb}^\ast$, so that subtracting
$A_1^{\rm IC}$ in Eq.~(\ref{ampimkp}) suppresses the strong
phase relative to the amplitude parameterized by $V_{us} V_{ub}^\ast E_1$
and hence suppresses $A_{\rm CP}$. The numerical sign associated
with $A_{\rm IC}$ is unclear; our remarks assume $B>0$
in Eq.~(\ref{ICestimate}) for definiteness. 
A suppressed $A_{\rm CP}$, as can be realized through the
inclusion of IC, is in better agreement with recent 
measurements, note 
$A_{\rm CP}(K^+\pi^-) = -0.07 \pm 0.08\, 
\hbox{(stat.)} \pm 0.02 \,\hbox{(sys.)}$~\cite{Aubert:2001qj}.
With the uncertainties from the IC contribution, we are compelled
to echo the conclusion of Ref.~\cite{Ciuchini:2001gv}: with the
present theoretical and experimental errors, it is not possible
to extract the value of $\gamma$ from these decays.

\section{Summary}

The role of non-valence components in the hadrons' light-cone
wavefunctions, coupled with the hierarchical structure of
the CKM matrix, offers new perspective on B decays. The effects
can be striking in decays for which the tree-level
contributions are Cabibbo-suppressed, so that
we have focused on the role of IC in the B meson
in mediating $B\to \pi K$ and
$B\to \rho K$ decays. We have shown that
such contributions  cannot be distinguished from
the ``charming'' penguin contributions of Ciuchini
et al.~\cite{Ciuchini:1997hb,Ciuchini:1998rj,Ciuchini:2001gv},
demonstrating that charming penguins need not be penguins at all.
The intrinsic charm
effect, although quantitatively elusive, is
sufficiently significant, on general grounds,
to impact the
$B\to \pi K$ and $B\to \rho K$ branching ratios, and the associated
partial rate
asymmetries, in a sizeable manner, though the largest
variations are seen in $A_{\rm CP}(K^+\pi^-)$ and
$A_{\rm CP}(K^+\pi^0)$.
Thus we must agree with the discouraging conclusions
of Ciuchini et al.~\cite{Ciuchini:2001gv}: the unestimated
contributions to $B\to \pi K$ decay are of sufficient size
that these modes cannot be used to estimate $\gamma$, or, better,
that any discrepancy between $\gamma$ determined in a fit of
the $B\to \pi K$ branching ratios and that of an unitarity
triangle fit is not yet meaningful.

The presence of IC in the B-meson touches many different
aspects of B physics. It can mediate decays which are 
suppressed in valence approximation, such as
$B\to J/\psi D\pi$~\cite{Chang:2001iy} and $B\to \chi_{0,2} K$.
It offers a pathway by
which the semi-leptonic branching ratio,
$B_{\rm sl}$, can be suppressed without exacerbating the charm
counting problem. It could potentially help explain the lifetime
difference between the $B$ meson and the $\Lambda_b$ baryon.

At the present stage of knowledge of non-perturbative QCD wavefunctions,
we only have a qualitative picture of
the structure of
non-valence components of the hadron's light-cone wave functions.
Eventually lattice gauge theory or methods such as DLCQ
may provide quantitative results.
It is very important to
re-measure the charm structure function of the proton
at large $x_{bj}$ to confirm or disprove the evidence for intrinsic
charm indicated by the
EMC measurements~\cite{Hoffmann:1983ah,Harris:1996jx,Steffens:1999hx}.
As emphasized by Chang and Hou\cite{Chang:2001iy}, 
the decay of the $B$ and $\Upsilon$ into
charmed final states may provide the best phenomenological constraints
on Fock states of the $B$ containing intrinsic charm.
We have emphasized here the impact of such Fock states
on exclusive decays to light hadrons.

Additional insight into the magnitude of IC and/or IS contributions
in the $B$ may  also be obtained
from decays in which all four quarks in the weak transition
differ, so that the penguin contributions which confound
the clear identification of non-valence effects in $B\to (\pi,\rho) K$
decays are absent.
For example, the decay $\bar B^0 \to K^- D^+$ proceeds at the
quark-level via $b \to c s \bar u$. In
the valence approximation
$W^-$ emission yields the only possible decay
topology. The $K^-$ meson forms from the decay of the $W^-$
to a color-singlet $s\bar u$, so that
the color transparency property of QCD suggests that
a factorization picture should be an excellent
description of this process. However additional decay
topologies are possible once non-valence degrees of freedom
are considered. For example, given intrinsic charm Fock states
in the B-meson, the $b$ and $\bar c$ can annihilate
and decay to $s \bar u$ and form a $K^-$ meson. The
remaining $c$ quark and spectator $\bar d$ form the $D^+$
meson. Additionally, using
IS in the B-meson, the $b$ and $\bar s$ can undergo
$W^-$ exchange to yield a $c \bar u$ final state.
These quarks can rearrange with the remaining $s$ quark and
spectator $\bar d$ to form a $D^+ K^-$ final state.
Thus the IC/IS-mediated contributions associated with the same
CKM factors and short-distance operators as the
conventional contributions 
could repair any quantitative deficiencies 
in the usual treatment of these decays.
However, since the IC/IS contributions in  such cases
is neither Cabibbo nor Wilson coefficient-enhanced relative
to the dominant contribution,
their presence may be difficult to confirm.

\acknowledgements
We are grateful to M. Beneke, G. Bonvicini, S. Dong,
I. Dunietz, G. Hiller, W.-S. Hou, P. Hoyer, D.-S. Hwang, A. Kagan, M.
Karliner, H.-N. Li,  H. Quinn, J. Tandean, and L. Wolfenstein for helpful
comments and discussions. S.G. thanks the SLAC Theory Group for
its hospitality during
the completion of this work.

%figures here

\vspace{3cm}

\vfill
\eject

% tables here

%%%%%%%%%%%%%%%%%%%%%%%%%%%%%%%%%%%%%%%%%%%%%%%%%%%%%%%%%%%%%%%%%%%
{\tabcolsep=0.2cm
\begin{table}[t]
\centerline{\parbox{14cm}{\caption{\label{tab:data}
Empirical branching ratios in $B\to \pi K$ decay
in units of $10^{-6}$ from the CLEO\protect\cite{CLEO00},
BaBar\protect\cite{BaBar01}, and Belle\protect\cite{Belle01}
experiments.
The symbol ${\bar {\cal B}}$ denotes the CP-averaged
branching ratio.
In constructing the averages, the statistical
and systematic errors for each experiment were added in quadrature.
}}}
\begin{center}
\begin{tabular}{|c|c|c|c|c|}
\hline
Mode & CLEO \protect\cite{CLEO00} & BaBar \protect\cite{BaBar01}
 & Belle \protect\cite{Belle01} & Average\hspace{0.15cm} \\
\hline
& & & & \\
${\bar {\cal B}}({\bar B}^0\to \pi^+ K^- )$ & $17.2_{\,-2.4}^{\,+2.5}\pm 1.2$
 & $16.7\pm 1.6 \pm 1.3$
 & $19.3_{\,-3.2\,-0.6}^{\,+3.4\,+1.5}$
 & $17.3\pm 1.5\quad$ \\
& & & & \\
${\bar {\cal B}}(B^-\to \pi^0 K^-)$ & $11.6_{\,-2.7\,-1.3}^{\,+3.0\,+1.4}$
 & $10.8_{\,-1.9}^{\,+2.1}\pm 1.0$
 & $16.3_{\,-3.3\,-1.8}^{\,+3.5\,+1.6}$
 & $12.1^{\, +1.7}_{\, -1.6}\quad$ \\
& & & & \\
${\bar {\cal B}}(B^-\to \pi^- {\bar K}^0)$ & $18.2_{\,-4.0}^{\,+4.6}\pm 1.6$
 & $18.2_{\,-3.0}^{\,+3.3}\pm 2.0$
 & $13.7^{+5.7 +1.9}_{-4.8 -1.8}$
 & $17.3^{\, +2.7}_{\, -2.4}\quad$ \\
& & & & \\
${\bar {\cal B}}({\bar B}^0\to\pi^0 {\bar K}^0)$ &
$14.6_{\,-5.1\,-3.3}^{\,+5.9\,+2.4}$
 & $8.2_{\,-2.7}^{\,+3.1}\pm 1.2$
 & $16.0^{+7.2 +2.5}_{-5.9 -2.7}$
 & $10.4^{\, +2.8}_{\, -2.5}\quad$ \\
& & & & \\
\hline%\hline
\end{tabular}
\end{center}
\end{table}}
%%%%%%%%%%%%%%%%%%%%%%%%%%%%%%%%%%%%%%%%%%%%%%%%%%%%%%%%%%%%%%%%%%%

%%%%%%%%%%%%%%%%%%%%%%%%%%%%%%%%%%%%%%%%%%%%%%%%%%%%%%%%%%%%%%%%%%%
{\tabcolsep=0.2cm
\begin{table}[t]
\centerline{\parbox{14cm}{\caption{\label{tab:IC}
Likely IC configurations in the ${\bar B}$ meson, where $q\in (u,d)$.
The $\ket{b c \bar{c} \bar{q}}$ light-cone wave function
will be maximized for states of minimal invariant mass, so
that the configurations with the lowest masses are the most likely.
The masses are assessed using the lowest-lying meson masses
in each $(q_1\bar{q_2})_{J^P}$ channel as described in the text.
The mass increment associated with the relative motion
of the two $q_1\bar{q_2}$ states has not been assessed, though
it likely acts to increase the estimated mass in the $l=1$
states.
}}}
\begin{center}
\begin{tabular}{|l|l|}
\hline
Configuration & Estimated Mass\\
\hline
$(b \bar{q})_{1^-}\; (c \bar{c})_{1^-}$ in a relative $l=1$ state
$\quad\quad\quad\quad\quad\quad$
&
$\rapp 8.4 \quad\quad$
$\quad\quad\quad\quad\quad\quad$
\\
& \\
$(b \bar{q})_{0^-}\; (c \bar{c})_{1^-}$ in a relative $l=1$ state &
$\rapp 8.4\quad\quad $
\\
$(b \bar{q})_{0^-}\; (c \bar{c})_{0^+}$ in a relative $l=0$ state &
$8.7 \quad\quad$
\\
& \\
$(b \bar{q})_{1^-}\; (c \bar{c})_{0^-}$ in a relative $l=1$ state &
$\rapp 8.3 \quad\quad$
\\
$(b \bar{q})_{0^+}\; (c \bar{c})_{0^-}$ in a relative $l=0$ state &
$8.7 \quad\quad$
\\
& \\
$(b \bar{c})_{1^-}\; (c \bar{q})_{1^-}$ in a relative $l=1$ state &
$\rapp 8.4 \quad\quad$
\\
& \\
$(b \bar{c})_{0^-}\; (c \bar{q})_{1^-}$ in a relative $l=1$ state &
$\rapp 8.3 \quad\quad$
\\
$(b \bar{c})_{0^-}\; (c \bar{q})_{0^+}$ in a relative $l=0$ state &
$8.7 \quad\quad$
\\
& \\
$(b \bar{c})_{1^-}\; (c \bar{q})_{0^-}$ in a relative $l=1$ state &
$\rapp 8.2 \quad\quad$
\\
$(b \bar{c})_{0^+}\; (c \bar{q})_{0^-}$ in a relative $l=0$ state &
$8.6 \quad\quad$
\\
\hline
\end{tabular}
\end{center}
\end{table}}


\begin{references}
\bibitem{kinem}
We define $p^\pm \equiv (p^0 \pm p^3)/\sqrt{2}$ as per
Ref.~\cite{lks} and
write $p=(p^+,p^-,\mathbf{p_\perp})$ where $\mathbf{p_\perp}=(p^1,p^2)$.
We work in the B meson rest frame, so
that $p_B=(m_B/\sqrt{2})(1,1,0_\perp)$.
\bibitem{details2}
The sign associated with the
$F_e^P$, $M_e^P$, $F_a^P$, and $M_a^P$ terms in $B\to \pi^0 K^0$
decay and its CP-conjugate, that is, in Eqs.(25) and (26) of
Ref.~\cite{lks} is in error.
\end{references}
\end{document}